\documentclass[sigconf]{acmart}

\usepackage{epsfig}
\usepackage{graphicx}
\usepackage{amsmath}
\usepackage{helvet}  %Required
\usepackage{courier}  %Required
\usepackage{url}  %Required
\usepackage{xcolor}
\usepackage{multirow}
\usepackage{subcaption}
\usepackage{array}
\usepackage{algorithm2e}
\usepackage{tabularx}
\usepackage{hyperref}

\usepackage{color, colortbl}
\definecolor{mygray}{gray}{0.9}

\newcolumntype{P}[1]{>{\centering\arraybackslash}p{#1}}

\newcommand{\blfootnote}[1]{%
  \begingroup
  \renewcommand\thefootnote{}\footnote{#1}%
  \addtocounter{footnote}{-1}%
  \endgroup
}

\def\BibTeX{{\rm B\kern-.05em{\sc i\kern-.025em b}\kern-.08emT\kern-.1667em\lower.7ex\hbox{E}\kern-.125emX}}
    
\copyrightyear{2024}
\acmYear{2024}
\setcopyright{rightsretained}
\acmConference[ICMR '24]{Proceedings of the 2024 International Conference on Multimedia Retrieval}{June 10--14, 2024}{Phuket, Thailand}
\acmBooktitle{Proceedings of the 2024 International Conference on Multimedia Retrieval (ICMR '24), June 10--14, 2024, Phuket, Thailand}\acmDOI{10.1145/3652583.3658032}
\acmISBN{979-8-4007-0619-6/24/06}

\begin{document}
\acmSubmissionID{392}
%
% The "title" command has an optional parameter, allowing the author to define a "short title" to be used in page headers.
% \title{A Novel Multimedia-based method for Assisting Ophthalmologists to Diagnose Retinal Diseases}
% \title{A Novel Multimedia-based method for Ophthalmologists to Diagnose Retinal Diseases}
% \title{Query-based Video Summarization with Self-supervision and Contextualized Word Representations}
% \title{Query-based Video Summarization with Self-supervision}
% \title[Conversational Image Search Meets Interactive Multimodal Learning]{Conversational Image Search Meets Interactive Multimodal Learning: Teaching a New Dog Old Tricks}
\title[Enhancing Interactive Image Retrieval With Query Rewriting]{Enhancing Interactive Image Retrieval With Query Rewriting Using Large Language Models and Vision Language Models}

% \title{relevance feedback-based Query Expansion for Multi-turn Image Retrieval}
% \title{GPT2MVS: Generative Pretrained Transformer 2 with hierarchical attention for Multi-modal Video Summarization}

% \author{Hongyi Zhu$^*$, Jia-Hong Huang$^*$, Stevan Rudinac, Evangelos Kanoulas}
% \affiliation{
%   \institution{University of Amsterdam, Netherlands \\
% %   \\ $^*$Work done during an internship at BBC Research and Development, London, UK.
% % {\tt\small waynewu@gatech.edu, j.huang@uva.nl, joee624@g.ucla.edu / Jachie.Lin@mediatek.com, m.worring@uva.nl}\\
% {\tt\small $*$ equal contribution}
%   }
% %   \city{Amsterdam Netherlands}
% %   \country{Netherlands}
%   }
% \email{h.zhu@uva.nl, j.huang@uva.nl, s.rudinac@uva.nl, E.Kanoulas@uva.nl}

\author{Hongyi Zhu$^*$}
\affiliation{
  \institution{University of Amsterdam}
  \city{Amsterdam}
  \country{The Netherlands}}
\email{h.zhu@uva.nl}

\author{Jia-Hong Huang$^*$}
\affiliation{
  \institution{University of Amsterdam}
  \city{Amsterdam}
  \country{The Netherlands}}
\email{j.huang@uva.nl}

% \author{Jia-Hong Huang$^*$}
% \affiliation{\institution{\small{University of Amsterdam, Netherlands} \\ \tt{j.huang@uva.nl}}}

% \author{Stevan Rudinac}
% \affiliation{\institution{\small{University of Amsterdam, Netherlands} \\ \tt{s.rudinac@uva.nl}}}

% \author{Evangelos Kanoulas}
% \affiliation{\institution{\small{University of Amsterdam, Netherlands} \\ \tt{E.Kanoulas@uva.nl}}}
% \author{Hongyi Zhu}
% \authornote{Both authors contributed equally to this research.}
% \email{h.zhu@uva.nl}
% \orcid{1234-5678-9012}
% \author{Jia-Hong Huang}
% \authornotemark[1]
% \email{j.huang@uva.nl}
% \affiliation{%
%   \institution{University of Amsterdam}
%   \city{Amsterdam}
%   \country{The Netherlands}
% }

\author{Stevan Rudinac}
\affiliation{%
  \institution{University of Amsterdam}
  \city{Amsterdam}
  \country{The Netherlands}}
\email{s.rudinac@uva.nl}

\author{Evangelos Kanoulas}
\affiliation{%
  \institution{University of Amsterdam}
  \city{Amsterdam}
  \country{The Netherlands}
}
\email{e.kanoulas@uva.nl}

%
% The abstract is a short summary of the work to be presented in the article.
\begin{abstract}
% Jiahong's version
Image search stands as a pivotal task in multimedia and computer vision, finding applications across diverse domains, ranging from internet search to medical diagnostics. Conventional image search systems operate by accepting textual or visual queries, retrieving the top-relevant candidate results from the database. However, prevalent methods often rely on single-turn procedures, introducing potential inaccuracies and limited recall. These methods also face the challenges, such as vocabulary mismatch and the semantic gap, constraining their overall effectiveness. To address these issues, we propose an interactive image retrieval system capable of refining queries based on user relevance feedback in a multi-turn setting. This system incorporates a vision language model (VLM) based image captioner to enhance the quality of text-based queries, resulting in more informative queries with each iteration. Moreover, we introduce a large language model (LLM) based denoiser to refine text-based query expansions, mitigating inaccuracies in image descriptions generated by captioning models. To evaluate our system, we curate a new dataset by adapting the MSR-VTT video retrieval dataset to the image retrieval task, offering multiple relevant ground truth images for each query. Through comprehensive experiments, we validate the effectiveness of our proposed system against baseline methods, achieving state-of-the-art performance with a notable 10\% improvement in terms of recall. Our contributions encompass the development of an innovative interactive image retrieval system, the integration of an LLM-based denoiser, the curation of a meticulously designed evaluation dataset, and thorough experimental validation. 

\end{abstract}




%
% A "teaser" image appears between the author and affiliation information and the body 
% of the document, and typically spans the page. 

% \begin{teaserfigure}
%   \includegraphics[width=\textwidth]{model_flowchart-AAAI_new.pdf}
%   \caption{
%     (a) is an existing traditional medical treatment procedure for retinal diseases. Please refer to the INTRODUCTION section for more details. In (b), we propose this multimedia-based method to improve (a) based on the point-of-care (POC) ~\cite{pai2012point} concept. In the proposed method, it mainly contains three modules, including DNN-based, DNN visual explanation, and multimedia visualization modules. Please refer to the METHOD section for the detail explanation. DNN indicates Deep Neural Networks.
%     }
% %   \Description{Enjoying the baseball game from the third-base seats. Ichiro Suzuki preparing to bat.}
%   \label{fig:figure1}
% \end{teaserfigure}

%
% This command processes the author and affiliation and title information and builds
% the first part of the formatted document.

\maketitle

\blfootnote{$*$ Equal contribution.}
\vspace{-0.5cm}
\section{Introduction}
% Jiahong's version
Image search is a fundamental task in multimedia and computer vision, with applications spanning internet search~\cite{noh2017large}, e-commerce product search~\cite{li2021embedding}, and medical research~\cite{hu2022x,pmlr-v182-zhang22a}. In a standard image-search system, a textual or visual query is provided as input, and the system computes similarity with images in the database to retrieve the top candidates. This task, also referred to as text-to-image and image-to-image retrieval~\cite{Suzuki2023TexttoImageFR,Mller2020MedicalIR,Zhou2017RecentAI}, plays a crucial role in various domains by facilitating efficient and effective access to visual information.

Over the years, a number of methods have been proposed to address the retrieval task~\cite{lu2022cots,sun2021lightningdot,jia2021scaling,lu2021visualsparta}. Existing approaches typically involve a single turn of the retrieval procedure: providing a text-based query to the retrieval model and directly obtaining the final retrieval results. However, this approach can be limited by challenges like vocabulary mismatch (e.g., different wordings for the same concepts) and the semantic gap (e.g., difficulty bridging the gap between text and the information it represents) ~\cite{895972}. 

Traditional information retrieval tackles these challenges using i.a., Pseudo-Relevance Feedback (PRF) ~\cite{Amati2002ProbabilisticMO, Robertson1991OnTS, Rocchio1971RelevanceFI, Zhai2001ModelbasedFI}. PRF operates under the assumption that some initially retrieved documents, despite keyword mismatches, are relevant to the user's intent (referred to as ``pseudo-relevant''). It then extracts terms from these documents to enrich the original query, enhancing its representativeness and potentially improving recall (the ability to find all relevant documents). This method has demonstrated success, even in multimedia retrieval tasks ~\cite{10.1145/1991996.1992047, rudinac2012leveraging}. 
In contrast to text document retrieval, image search encounters unique challenges. The disparity in modalities between the query and images precludes direct term extraction from retrieved images. To address this, a common strategy in applying feedback methods to image search is the vector space method ~\cite{Rocchio1971RelevanceFI, Rui1999ANR, Zahlka2018BlackthornLI, Khan2020InteractiveLF, 8843882}. This method iteratively adjusts the query vector based on positive and negative feedback, aiming to align it with relevant images and away from irrelevant ones within the multidimensional vector space. However, this approach is highly sensitive to retrieval results. Substantial modifications to the original query vector may result in the loss of crucial semantic content, potentially compromising search effectiveness.

\begin{figure*}[ht!]
  \includegraphics[width=\textwidth]{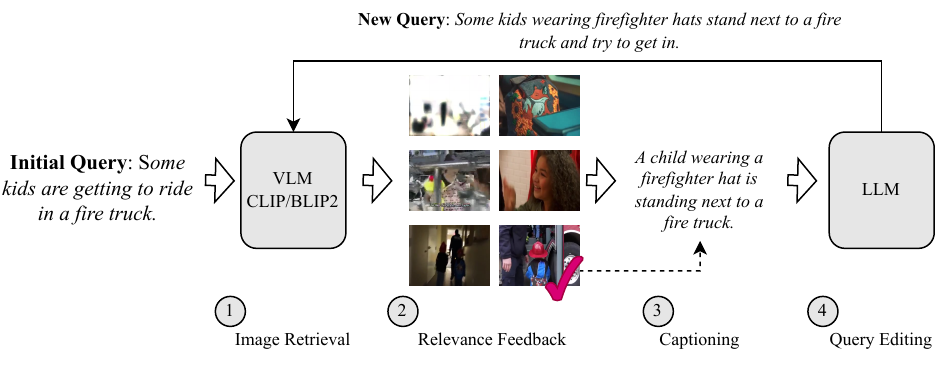}
  \vspace{-1.4cm}
  \caption{Flowchart of the proposed multi-turn interactive image retrieval approach based on relevance feedback. 
  This diagram illustrates an example of the initial interaction followed by query expansion and query refinement. 
  }
  \label{fig:figure2}
  \vspace{-0.2cm}
\end{figure*}

In response to the previously identified issues, we propose an interactive image retrieval system capable of presenting candidate images for a user query while continuously refining the query based on user relevance feedback in a multi-turn setting. Figure~\ref{fig:figure2} shows the flowchart of the proposed system. This method incorporates an image captioning model to augment the quality of the input text-based query in the natural language space. With each iteration of our retrieval approach, the query undergoes updates or expansions to generate a more informative text query.

However, in our proposed method, we have observed that certain image descriptions generated by an image captioning model might be inaccurate or lack specificity. When multiple image captions convey similar semantic meanings containing redundant information, it could mislead the retrieval model. In such scenarios, these descriptions introduce noise into the text-based query for the subsequent turn, potentially compromising the quality of subsequent queries. Consequently, the performance of the retrieval model might not improve or even decline at this stage. 
% referring to Section \ref{sec:result}.
To address this challenge, we propose the integration of an LLM-based denoiser. This denoiser is designed to refine the text-based query expansion before it is forwarded to the retrieval model for the next iteration.

Current single-turn image retrieval methods are primarily assessed on datasets like MSCOCO~\cite{Lin2014MicrosoftCC} and Flickr30k~\cite{Plummer2015Flickr30kEC}, designed for tasks such as object detection, segmentation, key-point detection, and image captioning. These datasets offer only one relevant image for each textual query, diverging from typical image retrieval datasets and posing limitations for our evaluation needs.
To effectively evaluate our proposed interactive image retrieval system, we curate a new dataset by adapting the video retrieval dataset MSR-VTT~\cite{Xu2016MSRVTTAL} to fit the image retrieval task. This modified dataset provides multiple relevant ground truth images for each query and is meticulously labeled by humans, featuring numerous ambiguous textual queries reflective of real-world application scenarios.

  In our experiments, we validate the effectiveness of the proposed interactive image retrieval system alongside baseline methods, including single-turn image retrieval models and vector space based relevance feedback models, using the newly constructed dataset. The experimental findings demonstrate that our proposed system outperforms the baselines by 10\% in terms of recall and achieves state-of-the-art performance.

% \vspace{+0.1cm}
The main contributions of this work are summarized as follows:
% \vspace{-0.2cm}

\begin{itemize}
    \item \textbf{Innovative Interactive Image Retrieval System:} Introducing a pioneering interactive image retrieval system that overcomes the limitations of existing single-turn methods. This system enables multi-turn interactions and continuous query refinement based on user relevance feedback. The incorporation of an image captioning model enhances the quality of text-based queries in natural language space, providing progressively informative queries with each iteration of the retrieval approach.

    \item \textbf{LLM-based Denoiser Implementation:} Proposing the integration of an LLM-based denoiser to refine text-based query expansions. This addresses inaccuracies and enhances specificity in image descriptions generated by image captioning models, resulting in improved query quality and overall retrieval performance.
    
    \item \textbf{Carefully Curated Evaluation Dataset:} Curating a meticulously designed dataset for evaluation by adapting the MSR-VTT video retrieval dataset to suit the image retrieval task. This dataset offers multiple relevant ground truth images for each query, addressing limitations present in datasets like MSCOCO and Flickr30k. It features human-labeled ambiguous textual queries that mirror real-world scenarios.

    \item \textbf{Thorough Experimental Validation:} Conducting comprehensive experiments to validate the effectiveness of the proposed interactive image retrieval system. The system is evaluated against baseline methods, which include single-turn image retrieval models and vector space-based relevance feedback models, utilizing the newly constructed dataset. The experimental results showcase a notable 10\% improvement in terms of recall after 6 interaction turns over the baselines, achieving state-of-the-art performance.
    
\end{itemize}

\vspace{-0.2cm}
\section{Related Work}
In this section, we present related works on image retrieval, interactive information retrieval methods, and query updating.
% \vspace{0.1cm}
% \noindent\textbf{2.1 Image Retrieval}
\subsection{Image Retrieval}

Image retrieval has emerged as a focal point of interest within the computer vision and information retrieval communities, finding widespread applications in various domains including e-commerce product search~\cite{li2021embedding}, face recognition~\cite{Schroff2015FaceNetAU,Parkhi2015DeepFR}, image geolocalization~\cite{Hays2008IM2GPSEG}, and medical image research~\cite{hu2022x,pmlr-v182-zhang22a}. 
One particularly intriguing aspect is cross-modal image retrieval, where queries and retrieval objects belong to different modalities. Examples include text-to-image retrieval~\cite{wang2015unsupervised, levy2024chatting}, cross-view image retrieval~\cite{Lin2015LearningDR}, and event detection~\cite{Jiang2015BridgingTU, wang2024prototype}. 
\\
Traditional methods for text-to-image retrieval have typically relied on convolutional neural networks (CNNs) as encoders to independently represent images and textual content \cite{Dong2014LearningAD,Radenovic2017FineTuningCI,noh2017large}. However, recent years have witnessed a surge in the adoption of transformer-based models and large-scale language-image pre-training, resulting in significant advancements \cite{radford2019language,lu2022cots,Lu2019ViLBERTPT}. These models have achieved state-of-the-art performance across various text-to-image benchmark tasks. Nevertheless, existing methods are primarily tailored for single-image retrieval and are evaluated on datasets like MSCOCO \cite{Lin2014MicrosoftCC} and Flickr30k \cite{Plummer2015Flickr30kEC}, which provide only one relevant ground truth image per textual query. The potential applications of these large-scale pre-trained VLMs in interactive image search tasks remain largely unexplored.

In this study, we focus on text-to-image retrieval, albeit within a human-machine interactive setting, aiming to explore the effectiveness of such models in facilitating interactive image search.

\subsection{Interactive Information Retrieval}
% \vspace{0.1cm}
% \noindent\textbf{2.2 Interactive Information Retrieval}

Interactive search has been instrumental in facilitating efficient access to document collections \cite{Allan1996IncrementalRF,Iwayama2000RelevanceFW,Joachims1997APA,Lewis1996TrainingAF}, evolving into an indispensable tool for multimedia researchers, especially in the early stages of content-based image and video retrieval \cite{Huang2008ActiveLF,Rui1997ContentbasedIR}. User-computer interaction can manifest in various formats, including relative attributes \cite{Parikh2011RelativeA,Kovashka2012WhittleSearchIS}, direct attributes \cite{zhao2017hierarchical,Han2017AutomaticSF,Ak2018LearningAR}, attribute-like modification text \cite{Vo2018ComposingTA}, and natural language \cite{Guo2018DialogbasedII,Guo2019TheFI}. Among these, feedback techniques such as relevance feedback \cite{Rocchio1971RelevanceFI,Iwayama2000RelevanceFW}, pseudo-relevance feedback \cite{Attar1977LocalFI,Xu1996QueryEU}, and implicit feedback \cite{Shen2005ContextsensitiveIR} are commonly used. 
Interactive image search aims to integrate user feedback as an interactive signal for navigating visual search, with feedback techniques widely studied and generally proven effective in enhancing retrieval accuracy \cite{Rocchio1971RelevanceFI,Robertson1991OnTS,Attar1977LocalFI}.

Numerous approaches have been proposed to enhance interactive image retrieval by integrating user feedback into the search query. Early efforts in interactive information retrieval focused on query expansion techniques based on PRF \cite{Amati2002ProbabilisticMO,Robertson1991OnTS,Rocchio1971RelevanceFI,Zhai2001ModelbasedFI}. In PRF-based methods, a set of retrieved documents from the initial query is treated as pseudo-relevant. The initial query is then expanded using the top-$k$ pseudo-relevant documents, and this expanded query is used for subsequent retrieval turns. PRF-based approaches offer practical advantages as they do not rely on constructing domain-specific knowledge bases and exhibit versatility across diverse corpora \cite{Jagerman2023QueryEB}.
While query expansion through PRF has shown promise in improving the recall of the retrieval system, its effectiveness is constrained by the quality of the top-$k$ pseudo-relevant documents. Non-relevant results within the feedback set introduce noise that may negatively impact retrieval quality. Additionally, image-based documents cannot directly expand textual queries for image retrieval tasks.

Another approach in interactive image retrieval methods is based on the vector space model \cite{Wang2008ASO}. In this model, the query vector is directly adjusted based on user feedback. The Rocchio method \cite{Rocchio1971RelevanceFI} is a commonly employed feedback technique within vector space models. This method involves updating the query vector using both relevant and non-relevant documents.
While vector space models can be applied to interactive image retrieval, directly modifying the query vector can significantly alter the hidden semantic information contained in the high-dimensional vector space. Alternatively, other relevant feedback-based vector space models \cite{Kovashka2012WhittleSearchIS,mikolov2013distributed,Khan2020InteractiveLF,Zahlka2018BlackthornLI} employ classifiers, such as linear Support Vector Machines, for images based on user-relevant feedback. Unlike methods that modify the query vector, these approaches do not involve direct modifications to the query vector. However, the linear classifier trained on a small amount of data in each interaction turn may lack generalization.

In our work, we primarily focus on relevance feedback-based image retrieval, where users are presented with a ranked list of candidate images in each interaction turn and are asked to assess their relevance. Specifically, we leverage a robust pre-trained VLM to extract features from relevant images, convert them into textual descriptions, and refine the query in natural language.

\subsection{LLM-based Query Editing}
% \vspace{0.1cm}
% \noindent\textbf{2.3 LLM-based Query Editing}

Existing query expansion models rely on pseudo-relevance feedback to enhance the effectiveness of initial retrieval. However, these models encounter challenges when the initial results lack relevance. In contrast, the authors of ~\cite{mackie2023generative} introduce Generative Relevance Feedback (GRF). This innovative approach constructs probabilistic feedback models using long-form text generated from LLMs. The authors demonstrate in ~\cite{mackie2023generative} that GRF methods surpass the performance of traditional PRF methods.

In ~\cite{jagerman2023query}, the authors introduce a method to query expansion that capitalizes on the generative capabilities of LLMs. Diverging from traditional methods like PRF, which depend on retrieving a set of pseudo-relevant documents to expand queries, the authors of ~\cite{jagerman2023query} exploit the creative and generative potential of an LLMs while tapping into its intrinsic knowledge. The study encompasses a range of prompts, including zero-shot, few-shot, and Chain-of-Thought. Notably, the authors observe in ~\cite{jagerman2023query} that CoT prompts prove particularly effective for query expansion. These prompts guide the model to systematically break down queries, yielding an extensive array of terms closely linked to the original query. 

In ~\cite{mao2023search}, the authors introduce EdiRCS, a novel text editing-based conversational query rewriting model designed specifically for conversational search scenarios. EdiRCS adopts a non-autoregressive approach where most rewrite tokens are drawn directly from the ongoing dialogue, minimizing the need for additional token generation. This design choice enhances the efficiency of EdiRCS significantly. Notably, the learning process of EdiRCS is enriched with two search-oriented objectives: contrastive ranking augmentation and contextualization knowledge transfer. These objectives are instrumental in improving EdiRCS's ability to select and generate tokens that are highly relevant from a retrieval perspective. 

Understanding users' contextual search intent accurately is a significant challenge in conversational search, given the diverse and long-tailed nature of conversational search sessions. In ~\cite{mao2023large}, the authors introduce LLMs4CS, a straightforward yet effective prompting framework designed to harness the power of LLMs as text-based search intent interpreters for conversational search. Within this framework, the authors of ~\cite{mao2023large} explore three prompting methods to generate multiple query rewrites and hypothetical responses. In ~\cite{mao2023large}, the authors propose aggregating these outputs into an integrated representation capable of robustly capturing the user's true contextual search intent. 

Query rewriting is vital for enhancing conversational search by converting context-dependent user queries into standalone forms. In ~\cite{ye2023enhancing}, the authors advocate for employing LLMs as query rewriters to generate informative query rewrites with carefully crafted instructions. They establish four essential properties for well-formed rewrites and emphasize their integration into the instructions. In addition, they introduce the concept of rewrite editors for LLMs in a ``rewrite-then-edit'' process, particularly when initial query rewrites are available. Finally, they propose condensing the query rewriting capabilities of LLMs into smaller models to reduce latency in the rewriting process. Different from previous studies, in this paper, we jointly utilize the capability of both the VLM and the LLM to extract information from the multimodal interactive search context and gradually improve the search query.  

In this study, we present a relevance feedback-based interactive image retrieval system that integrates an image captioning model to enhance the quality of text-based queries in natural language, thereby generating increasingly informative queries with each iteration of the retrieval process.

%%%%%%%%%%%%%%%%%%%%%%%%%%%%%%%%%%%%%%%%%%%%%%%%%
%%%%%%%%%%%%      Methodology       %%%%%%%%%%%%%
%%%%%%%%%%%%%%%%%%%%%%%%%%%%%%%%%%%%%%%%%%%%%%%%%

% \begin{figure*}[ht!]
%   \includegraphics[width=\textwidth]{samples/single_turn_numbered_blocks.drawio.pdf}
%   \vspace{-0.6cm}
%   \caption{Flowchart of the proposed relevance feedback-based multi-turn interactive image retrieval. This graph also presents an example of the initial turn interaction and the subsequent query expansion and query edit. The whole system consists of four steps, it starts with the initial textual query from the user, then the Dense retriever encodes the query into a feature vector and computer the similarity with every image in the database. The Interaction module would present the top-$k$ similar images to the user (in our system is the artificial actor) and require the user to select the relevant images among them. We use the Vision Language Model captioner to generate the descriptions for the relevant images and expand them to the original query. The step is a query edit which would reformulate the expanded query and generate a more effective query. This process would be conducted repeatedly and keep updating the search query and rerank new unseen images.}
%   \label{fig:figure2}
%   \vspace{-0.2cm}
% \end{figure*}

\vspace{-0.1cm}
\section{Methodology}
In this section, we present our innovative interactive image retrieval system and discuss the accompanying LLM-based query reformulation method. The proposed image retrieval approach unfolds as an iterative process with three key steps. First, the Image Retrieval step utilizes a pre-trained cross-modal dense retrieval model. Second, an Artificial Actor, following Zahálka et al. \cite{Zahlka2015AnalyticQE}, emulates a real user by providing relevance feedback and evaluating the retrieval results. Third, the Query Expansion step employs a VLM to expand the query based on the relevance feedback from the Artificial Actor~\cite{Zahlka2015AnalyticQE}. The workflow of the proposed approach operates in a multi-turn setting, encompassing image retrieval, relevance feedback, and query expansion. Subsequently, we introduce our approach that employs an LLM-based method to denoise expanded queries, thereby enhancing the quality of the expanded queries.
\subsection{Image Retrieval}
  
In this study, we concentrate on the task of interactive text-to-image retrieval. Given a textual query $q$ and an image collection $\mathcal{I}$, the image retrieval system returns a ranked list of images $\mathcal{L}$. Here, $L_{i}$ represents the $i$-th ranked image in the list. The system's objective is to retrieve as many relevant images as possible from the top-$k$ ranked images $\mathcal{L}_{k}$. We focus solely on the top-$k$ retrieval results, considering that users typically review only the top search results \cite{lewandowski2008retrieval, Ricci2011IntroductionTR}.
We employ a pre-trained VLM like CLIP \cite{radford2019language} to extract the visual embedding $f_{v}$ from the image dataset. The textual query is embedded as $f_{t}$. During retrieval, we extract $f_{t}$ for the given query text and compute the cosine similarity with the visual vector $f_{v}$ of all images in the candidate pool to identify the top-$k$ similar images.

Our research aims to explore how to iteratively use the relevance feedback and the top-$k$ retrieved images $\mathcal{L}_{k}$ to rerank the next $k$ unseen images in the original ranked list: $\mathcal{U}=\left\{L_{nk+1}, \ldots, L_{nk+k}\right\}$. Here, $n$ denotes the number of interactive turns, and $\mathcal{U}$ represents the unseen images to be ranked.
%  \textcolor{red}{The subsections of your method should follow your method flowchart.}
% \sr{Yes, indeed, I agree with Evangelos - I labeled different blocks in the approach pipeline, so the sub-sections could have the same index.}

% \vspace{0.1cm}
% \noindent\textbf{3.2 Relevance Feedback}
\subsection{Relevance Feedback}

In this study, we employ the concept of an Artificial Actor \cite{Zahlka2015AnalyticQE} to assess the relevance of the feedback images. The artificial actor is a computational agent capable of interacting with the evaluation method and simulating user behavior. The construction of artificial actors is based on three fundamental principles: analytic categories, evolving notions of relevance, and limited time. Analytic categorization refers to the task of classifying individual items into categories defined by the analyst (user). The actors adapt their categories of relevance over time, thereby modeling the dynamic nature of insight. The artificial actor aims to emulate the user's behavior during the analysis of a collection and the development of insight over time. In other words, the user's needs, intent, and the notion of relevance evolve over the course of the analysis.

In our task, we only use a simple version of the artificial actor. Given a textual query $q$ and retrieved top-$k$ images $\mathcal{L}_{k}$ from the feedback set, we use the ground truth label from the dataset to binary determine the relevance of each image $\mathcal{L}_{i}$ and randomly select $n$ relevant images to form the final relevant image set $\mathcal{L}_{r}^{q}$. The reason for this design is that the artificial actor in our experiment is not required to assign images to a category set and the only task is to determine whether an image is relevant or irrelevant to a given textual query. The random selection is to simulate the artificial actor that would change the notion of relevance over time. Moreover, the time efficiency of the retrieval system is also not our research topic, so this characteristic of the artificial actor is also not considered. 

We operate under the assumption that a real user can always accurately determine the relevance of a candidate image to the given query. Therefore, the use of ground truth labels from the dataset annotation, combined with a random sampling strategy, can effectively simulate the user's real behavior. Importantly, the ground truth of the dataset label remains unseen to the pre-trained dense retriever, thereby eliminating any potential issues related to information leakage.

% \vspace{0.1cm}
% \noindent\textbf{3.3 Query Expansion}
\subsection{Query Expansion}

We define the query expansion problem as follows: Given a textual query $q$, our goal is to generate an expanded query $q^{\prime}$ that incorporates additional information absent from the original query. Specifically, we investigate the application of state-of-the-art VLMs to generate captions for the images in the relevant image set $\mathcal{L}_{r}$. Formally, this process can be represented as:
\begin{equation}
q^{\prime}=\operatorname{Concat}\left(q, \operatorname{VLM}\left(\text{Prompt}_{\mathcal{L}_r}\right)\right)
\end{equation}
Where Concat means the string concatenation operation, $q$ is the original query, the VLM is a pre-trained VLM-based image captioner, and $\text{Prompt}_{\mathcal{L}_r}$ is the prompt based on the query. 
% These models are pre-trained on billion-level data of image-text pairs from the internet, offering unique representations with robust generalization and transfer capabilities.Current VLMs excel at integrating information from both modalities and have demonstrated impressive zero-shot performance for numerous downstream vision applications.

Image captioning is a fundamental task in multimedia modeling. Recent models predominantly rely on large-scale VLMs ~\cite{Desai2020VirTexLV, Sariyildiz2020LearningVR, huang2017vqabq, Li2021SupervisionEE}. Current VLMs excel at integrating information from both modalities and have demonstrated impressive zero-shot performance for numerous downstream vision applications. In this study, we employ the state-of-the-art VLM, InstructBLIP ~\cite{Dai2023InstructBLIPTG}, to generate captions for relevant images in a zero-shot setting. The instruction prompt used for image captioning is ``$\emph{<Image>~A short image caption:}$''. This prompt guides InstructBLIP to generate a concise image description sentence with fewer than 100 tokens. The rationale for generating short image descriptions is to circumvent the issue of hallucination ~\cite{Li2023EvaluatingOH, Hu2023CIEMCI, Zhai2023HallESwitchRA} in the VLM-based caption model. A brief image caption introduces less redundant information to the query, thereby enhancing the efficiency of the retrieval process.

\subsection{LLM-based Query Editing.}
Following each interactive image retrieval turn, a given textual query is concatenated with the captions of the images from the relevant image set $\mathcal{L}_{r}^{q}$, provided there is at least one relevant image among the top-$k$ retrieved images $\mathcal{L}_{k}$. Ideally, the length of the query increases with each interaction. However, an over-expanded query can negatively impact retrieval performance, as multiple captions with similar or repetitive content can mislead the retrieval model. Additionally, the text encoder of the retrieval model has a limit on input length, leading to the truncation of overlong queries. To address these issues, we explore the use of LLMs to denoise the redundant information in the expanded query and further enhance query quality. We assume that the original query $q$ is expanded in every interaction turn, resulting in the expanded query $\mathcal{C}^{t} = \left(q, q^1, \ldots, q^t \right)$ for the current turn $t$.

\subsubsection{Prompting Method.}
The prompt in our study comprises three components: \emph{instruction}, \emph{demonstration}, and \emph{input}. The \emph{instruction} defines the specific generation subtask for the query. The \emph{demonstration} provides in-context examples. The \emph{input} consists of the original query and the expanded image captions. Specifically, the \emph{input} is $\mathcal{C}^{t}$, i.e., the concatenation of the original query and all the captions within turn $t$. We investigate how LLMs can generate modifications of the expanded query $\mathcal{C}^{t}$. To this end, we design and explore three prompting methods to guide the LLMs in conducting three query-specific generation subtasks: Query Summary, Keywords Summary, and Chain of Thought (CoT) generation. Examples of prompts for these three subtasks and the corresponding results generated by the LLMs can be found in Table \ref{tab:prompt_example}.

\textbf{Query Summary.}
In this work, we leverage the robust automatic summarization capabilities of LLMs ~\cite{Kojima2022LargeLM, zhang2016summary, Wang2023ElementawareSW} to reformulate the expanded query in an in-context learning setting. With this approach, we treat LLMs as proficient query rewriters and prompt them to generate more concise and clear queries. Detailed templates for this prompting method and generation results can be found in the ``summary'' section of Table \ref{tab:prompt_example}. Despite its simplicity, as demonstrated in Section \ref{sec:result}, this straightforward prompting method has proven to deliver strong search performance, outperforming existing baselines.

\textbf{Keywords.}
Keywords form another crucial element within the query. The task of keyword extraction closely aligns with Query Facet Extraction. Implementing query facet extraction in a web search system can aid in refining and specifying the original query, exploring various subtopics, and diversifying the search result~\cite{Oren2006ExtendingFN, Samarinas2022RevisitingOD, Deveaud2014AccurateAE}. 
To leverage the keyword or facet contained in the query, we direct the LLMs to generate a list of significant words or phrases present in the expanded query. The ``keywords'' section of Table \ref{tab:prompt_example} provides the application details. In the subsequent retrieval turn, we concatenate the generated keywords with the prefix ``Video or image of'' and output the new query. For instance, ``video or image of kids, firefighter, hat,...''. This format closely resembles the prompt of the classification task of the CLIP model~\cite{Radford2021LearningTV}.

\textbf{CoT Summary.}
CoT~\cite{Wei2022ChainOT} or basic question answering~\cite{huang2017vqabq,huang2017robustness,huang2019novel,huang2017robustnessMS,huang2023improving,huang2019assessing} empower LLMs to break down a reasoning task into multiple intermediate steps, thereby enhancing the reasoning abilities of LLMs. Recent research~\cite{Wang2023ElementawareSW} has explored the use of CoT to guide LLMs in generating summaries in a step-by-step manner. This approach guides LLMs to extract the four most crucial elements such as \emph{Entity, Date, Event, and Result} from a standardized news text and subsequently generate a summary. In this study, we also examine whether using CoT reasoning can further enhance the quality of the query. However, the text we process is open-domain image captions, which lack the standard article structure found in news articles. Thus, we do not predefine the category of keywords that need to be extracted from the expanded query. 

More specifically, as illustrated in the ``CoT Summary'' section of Table \ref{tab:prompt_example}, we manually design a prompt that first guides LLMs to extract the keywords, topics, and taggings from the expanded query. Then, by using the extracted keywords, a new query is generated. The extraction of keywords from multiple relevant image captions can filter out duplicated content without losing key information. Regenerating the query based on the keywords can restore the semantic structure of the query sentence. As demonstrated in Section~\ref{sec:result}, our proposed CoT method significantly improves the quality of the query. 

\begin{table*}[t!]
\centering
\caption{A general depiction of the query editing process using LLMs. In the context of a video search dataset, a typical prompt follows this format: ``Given the video descriptions, generate $\left\{\textbf{subtask}\right\}$, $\left\{\textbf{demonstration}\right\}$, $\left\{\textbf{query}\right\}$'', where \textbf{demonstration} refers to the few-shot examples, which are not included in this table.
% A general illustration of the LLMs-based query editing.  For a video search dataset, a standard prompt would be in the format of this: ``Given the video description, generate $\left\{\textbf{subtask}\right\}$, $\left\{\textbf{demonstration}\right\}$, $\left\{\textbf{query}\right\}$'', where \textbf{demonstration} means the few-shot examples which are omitted in this table.
}
\vspace{-0.3cm}

\begin{tabularx}{\textwidth}{l|l}
\hline
\textbf{\begin{tabular}[c]{@{}l@{}} Original Query\\ Captions\end{tabular}} & \begin{tabular}[c]{@{}l@{}} \textit{Original Query:} A kid wearing a hat going inside the bus and 2 persons are standing behind them.\\ \textit{Generated Captions:} \\ A firefighter helps a young boy into the back of a fire truck. \\
A young boy wearing a firefighter hat is standing next to a fire truck.\end{tabular} \\ \hline
\textbf{Subtask:}                          & \textbf{Prompt+LLMs model output}                   \\
                                 &                              \\
Summary                          &         \begin{tabular}[c]{@{}l@{}} \textit{Prompt:} Given the video descriptions, generate the topic of this video. \\
\textit{Model output:} Children wearing hat boarding a fire truck, with the assistance of a firefighter \end{tabular}
                                                                    \\
                                 &                                  \\ 
Keywords                         &        \begin{tabular}[c]{@{}l@{}} \textit{Prompt:} Given the video descriptions, generate a list of video taggings \\
\textit{Model output:} Video of kids, bus, Firefighter, Fire truck, hat.               \end{tabular}                                                     \\
                                 &                         \\ 
CoT Summary               & \begin{tabular}[c]{@{}l@{}} \textit{Prompt:} Given the video descriptions, generate a list of keywords and a new description based on generated keywords. \\ \textit{Model output:} \\ keywords: kids, bus, Firefighter, Fire truck, hat\\ A group of kids as they board a firetruck, with the help of a friendly firefighter. ready for an exciting adventure!\end{tabular}         \\ \hline                                                        
\end{tabularx}%

\label{tab:prompt_example}
% \vspace{-0.2cm}
\end{table*}

\section{Experiments and Analysis}
In this section, we provide a detailed overview of the experimental settings. The evaluation of our proposed interactive image retrieval approach is conducted on our proposed modified version of the MSR-VTT video search dataset ~\cite{Xu2016MSRVTTAL}. Subsequently, we analyze the effectiveness of query expansion using relevant image captions and LLMs-based query editing. The code and models are released at \href{https://github.com/s04240051/Multimodal-Conversational-Retrieval-Sytem}{https://github.com/s04240051/Multimodal-Conversational-Retrieval-System.git}

% In this section, the experimental settings are described in detail. The proposed interactive image approach is evaluated on a modified version of the video search dataset MSR-VTT ~\cite{Xu2016MSRVTTAL}. Then, the effectiveness of query expansion with relevant image captions and LLMs-based query edits is analyzed.

% \vspace{0.1cm}
% \noindent\textbf{4.1 Dataset and Evaluation Matric}
\subsection{Dataset and Evaluation Matric}

% \sr{This is a point that needs immediate attention. Unless I am completely missing something, we perform evaluation on only one dataset, i.e. MSR-VTT. However, in Abstract we write the following ``Evaluation on two datasets demonstrates our method’s efficacy in improving recall compared to single-turn retrieval and
% vector space-based models.''. Please modify Abstract to reflect what has been done.}

\textbf{Dataset.} The MSR-VTT (Microsoft Research Video to Text) dataset \cite{Xu2016MSRVTTAL} is a comprehensive, large-scale collection curated for video description tasks. It comprises 7,180 videos, collected using a commercial video search engine by retrieving 118 videos per query for 257 popular queries. From this dataset, 10,000 video clips were selected, amounting to 41.2 hours, along with 200,000 clip-sentence pairs. These pairs, covering a wide range of visual content, serve as a valuable resource for advancing video-to-text methodologies and enhancing our understanding of video content through natural language. Each video clip is annotated with approximately 20 natural sentences by a team of 1,327 Amazon MTurk workers.

We have transformed the MSR-VTT dataset into a text-to-image or text-to-frame retrieval dataset, which provides multiple ground truth relevant images for a given query. For each clip, we extract one keyframe per second. Given that the length of each clip ranges between 10 and 20 seconds, we randomly select 10 frames as the ground truth relevant images. As each clip has 20 textual captions, they are treated separately as distinct data samples, each mapping to the same 10 ground truth relevant images in that clip. We utilize the entire clip set for the experiment, resulting in 200,000 data samples (queries) and 100,000 images in the entire collection.

In retrieval, the proposed system aims to retrieve as many relevant images as possible for a given query. Therefore, the use of video datasets enables the retrieval of multiple relevant images for a single query. Moreover, the frames distributed in a short clip generally contain the same scene and topic but from different camera perspectives. Each clip also has 20 unique captions that describe the clip in different ways or about different time slices. Each data sample consists of one textual query and 10 images. This setting allows the search query to closely resemble real human language, which is typically incomplete and ambiguous.

\textbf{Evaluation Metric.} In line with existing work~\cite{Borlund2003TheIE}, we employ accumulated recall metrics to evaluate the performance of our proposed multi-turn interactive image retrieval system. In each interaction turn, we display and rank 20 images, conducting a total of 6 turns. Images that have been ``seen'' in previous turns are not retrieved again. Given a query, the accumulated recall of the multi-turn search system is calculated as the number of relevant images found in all previous turns divided by the total number of ground truth relevant images for that query. For the entire dataset, the average accumulated recall $R^{T}$ of the system by turn $T$ is calculated as follows:
\begin{equation}
    R^{T} = \frac{1}{N} \sum_{i=1}^{N} \frac{\sum_{t=0}^{T} |\mathcal{L}_{ri}^{t}|}{n_{r}^{i}}
\end{equation}
Where $N$ is the number of data samples (queries), $|\mathcal{L}_{ri}^{t}|$ means the number of relevant images in turn $t$ for the data sample $i$,  ${n_{r}^{i}}$ means the number of ground truth relevant image of a given query $q_{i}$. For the single-turn image retrieval, we use Recall@20, Recall@40, Recall@60, Recall@80, Recall@100, and Recall@120 which is comparable to the multiturn image retrieval evaluation.

% \vspace{0.1cm}
% \noindent\textbf{4.2 Experimental Settings}
\subsection{Experimental Settings}

In our experiment, we employ the pre-trained CLIP model (``ViT-L/14@336px'')~\cite{Radford2021LearningTV} to encode both textual queries and images. Additionally, we use the BLIP-2~\cite{Li2023BLIP2BL} retriever model, which utilizes a pre-trained ``ViT-L/14@336px'' model to extract embeddings. For image captioning, we use the InstructeBLIP~\cite{Dai2023InstructBLIPTG} model, which leverages a pre-trained language model, vicuna-7b ~\cite{Zheng2023JudgingLW}, to generate captions and a ViT-g/14~\cite{Dosovitskiy2020AnII} model to extract features from the image. The prompt used for image captioning is ``<Image> A short image caption:'', which guides the model to generate a sentence of fewer than 100 tokens, excluding special symbols. The primary LLM used for query editing is the pre-trained LLama2-7b model~\cite{Touvron2023Llama2O}, which is instruct-tuned on a dialogue dataset. In each interaction turn, we only display the top-$k=20$ ranked images to the user (artificial actor) for relevance judgment. This approach is based on research~\cite{lewandowski2008retrieval} indicating that users typically focus only on the top search results during web searches. For image searches, users may have the patience to review more results, but 20 is generally the limit. Given a search session, we conduct only five turns of interactive searches, regardless of whether the system finds relevant images. This is because users typically lack the patience to provide more turns of relevance feedback, as suggested by Wang and Ai~\cite{Wang2021ControllingTR}. We expand the query with captions from at most two relevant images after the user's relevance judgment to improve the system's efficiency and prevent the query from containing too much redundant information. Regarding the procedure of interaction turns, turn 0 involves retrieval with the original query, turn 1 only conducts query expansion with image captions, and LLM-based query editing is applied only after turn 1.

% \vspace{0.1cm}
% \noindent\textbf{4.3 Comparison Methods}
\subsection{Comparison Methods}

We compare the proposed interactive image retrieval model to two types of models, the single-turn VLM, and the multi-turn vector space model. 

\textbf{CLIP~\cite{Radford2021LearningTV}}: This is a widely utilized Vision-Language Model (VLM)-based single-turn dense retriever. We employ the pre-trained CLIP model, which incorporates a ``ViT-L/14@336px'' image encoder and a Bert-based text encoder~\cite{Devlin2019BERTPO}.

\textbf{BLIP-2~\cite{Li2023BLIP2BL}}: This is the state-of-the-art VLM-based single-turn dense retriever. We utilize BLIP-2~\cite{Li2023BLIP2BL}, which features a ``ViT-L/14'' image encoder and the multimodal feature fusion module Q-Former. The latter is initialized with $Bert_{base}$~\cite{Devlin2019BERTPO} and pre-trained on datasets such as MSCOCO~\cite{Lin2014MicrosoftCC} and LAION400M~\cite{Schuhmann2021LAION400MOD}.

\textbf{Rocchio~\cite{Rocchio1971RelevanceFI}}: This is a commonly employed feedback method that modifies the query in the vector space. The concept involves updating a query vector with both relevant and non-relevant documents. The Rocchio method leverages the textual query embedding vectors and image embedding vectors encoded by the corresponding retriever model. We set $\alpha=1$, $\beta=0.75$, and $\gamma=0.15$.

% \vspace{0.1cm}
% \noindent\textbf{4.4 Main Results}

\subsection{Main Results}
\label{sec:result}
Our system incorporates two VLM-based dense retrieval models, namely CLIP and BLIP-2. As the retrievers form a crucial component of our system, it is essential to compare the performance of these two commonly used cross-modal retrievers, which differ in their structure and pre-training datasets. The impact of different LLMs is discussed in Section \ref{sec:ablation}. The overall performance comparisons of these models are presented in Table \ref{tab:table2}.  

\textbf{Effect of Interactive Search.} The implementation of an interactive search mechanism resulted in an enhancement of the single-turn image retrieval system's recall by over 10\%. This significant improvement underscores the marked advantage of our interactive search system in comparison to conventional methods.

\begin{figure}[t!]
   \includegraphics[width=\linewidth]{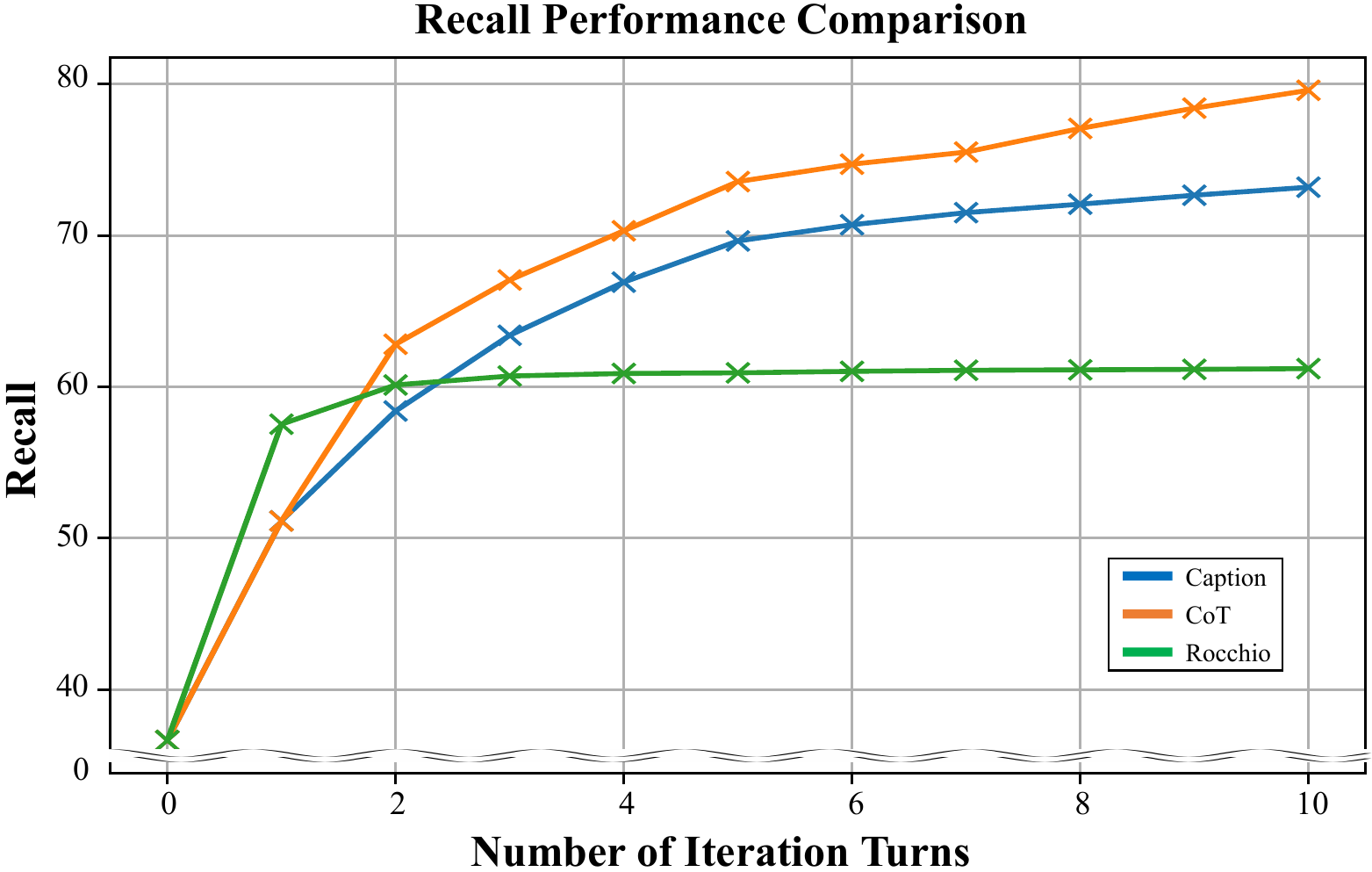}
  \vspace{-0.6cm}
  \caption{The trend in accumulated recall of the methods shows a steady increase with the number of interaction turns. However, the line representing the Rocchio method levels off after the second turn, indicating that no further relevant images are retrieved beyond this point. In contrast, the other two methods demonstrate the ability to consistently retrieve new relevant images across successive turns.
  % The trend of the methods' accumulated recall increases with the number of interactive turns. The Rocchio method's line ceases to incline, indicating that no additional relevant images are retrieved after the second turn. In contrast, the other two methods are capable of continuously retrieving new relevant images.
  }
  \label{fig:figure3}
  \vspace{-0.6cm}
\end{figure}

\textbf{Effect of Query Expansion.}
The column labeled “Captions only” in Table \ref{tab:table2} demonstrates that a simple query expansion using the captions of relevant images yields a robust search performance, surpassing other methods. This approach even exceeds the results of the in-context learning query summary and closely matches the performance of the keywords summary method, albeit slightly inferior to the CoT summary. Upon further comparison of these two methods over additional interaction turns, as depicted in Figure \ref{fig:figure3}, the rate of performance improvement for query expansion with image captions significantly decelerates after the fifth turn. In contrast, the CoT query summary continues to enhance its performance. This observation suggests that over-expansion of the original query with excessive image captions can degrade the quality of the query. The LLM-based query editing emerges as an effective denoising tool, capable of eliminating redundant information and refining the expanded query. Moreover, Figure \ref{fig:figure3} reveals that the Rocchio method achieves peak performance at turn 1 (the initial query expansion turn) but ceases to improve beyond turn 2. This outcome indicates that query expansion based on the vector space model is less effective compared to methods that perform query expansion within the natural language space.

\textbf{Effect of Different Retrievers.}
As indicated in Table~\ref{tab:table2}, the BLIP-2-based retriever exhibits a 1\% increase in recall compared to the CLIP-based retriever in the single-turn retrieval task. However, BLIP-2 does not offer a substantial benefit in the multi-turn retrieval task. Given that BLIP-2 comprises more parameters and has lower inference efficiency, it is advisable to utilize the CLIP-based retriever in subsequent experiments.

\textbf{Effect of Different Prompting Methods.} As evidenced in Table~\ref{tab:table2}, the interactive image retrieval system that utilizes the CoT summary surpasses all other prompting methods post the second turn. The integration of the CoT into the prompting methods typically enhances the search performance. This underscores the effectiveness of our CoT in steering the LLM toward an accurate comprehension of both the scene details in the individual keyframe and the plot specifics in the video clip. 

\textbf{Case Study.} In Figure \ref{fig:figure example}, we showcase some queries and search results from our most effective model. We present both successful and unsuccessful cases. These examples demonstrate that queries with a well-structured format and notable keywords, such as ‘firetruck’, tend to yield better retrieval results. However, the database may contain numerous negative cases that closely resemble the truly relevant image. As illustrated in the concluding case of Figure \ref{fig:figure example}, it becomes exceedingly arduous to differentiate between negative and positive images, given that they portray an identical scene from a television program.

\begin{table}[t!]
\centering
\caption{
Overall performance comparison. The column labeled as ``Single Turn'' shows the results of one-turn retrieval using CLIP and BLIP-2. For multi-turn retrieval, we only display the top-$k=20$ ranked images to the user (artificial actor) for relevance judgment in each turn. The best scores in each turn, except for turn 0 (where no query expansion or editing methods are applied), are underlined for clarity.
% The overall performance comparison. The column named ``Single Turn'' indicates the one-turn retrieval result of CLIP and BLIP-2. The column named ``Caption only'' means the result of experiments that only expand the query with the captions of the relevant images. The best scores in each turn except turn $0$ (no query expansion or query edit methods are used in turn $0$) are underlined.
}
\vspace{-0.3cm}

\scalebox{0.65}{
\begin{tabular}{|cccccccc|}
\hline
\multicolumn{1}{|c|}{}                                                         & \multicolumn{1}{c|}{\begin{tabular}[c]{@{}c@{}}Single\\ Turn\end{tabular}}                            & \multicolumn{1}{c|}{\begin{tabular}[c]{@{}c@{}}Multiple\\ turns \\ Recall\end{tabular}}                          & \multicolumn{1}{c|}{Rocchio}                             & \multicolumn{1}{c|}{\begin{tabular}[c]{@{}c@{}}Captions\\ Only\end{tabular}}     
            & \multicolumn{1}{c|}{\begin{tabular}[c]{@{}c@{}}With \\ Summary\end{tabular}}  
            & \multicolumn{1}{c|}{\begin{tabular}[c]{@{}c@{}}   Keywords \\ Summary \end{tabular}}                      & \begin{tabular}[c]{@{}c@{}}CoT \\ Summary\end{tabular}                               
            \\ \hline
\multicolumn{1}{|c|}{\begin{tabular}[c]{@{}c@{}}CLIP \\ Retriever\end{tabular}} & \multicolumn{1}{c|}{\begin{tabular}[c]{@{}l@{}}R@20: 36.64\\ R@40: 45.69\\ R@60: 51.01\\ R@80: 54.75\\ R@100:57.73\\ R@120:60.08\end{tabular}} & \multicolumn{1}{c|}{\begin{tabular}[c]{@{}l@{}}Turn 0\\ Turn 1\\ Turn 2\\ Turn 3\\ Turn 4\\ Turn 5\end{tabular}} & \multicolumn{1}{c|}{\begin{tabular}[c]{@{}l@{}}36.64\\ \underline{57.53}\\ 60.12\\ 60.71\\ 60.88\\ 60.92\end{tabular}} & \multicolumn{1}{c|}{\begin{tabular}[c]{@{}l@{}}36.64\\ 51.01\\ 58.18\\ 63.11\\ 66.58\\ 69.34\end{tabular}} & \multicolumn{1}{c|}{\begin{tabular}[c]{@{}l@{}}36.64\\ 51.16\\ 56.81\\ 60.68\\ 63.60\\ 65.86\end{tabular}} & \multicolumn{1}{c|}{\begin{tabular}[c]{@{}l@{}}36.64\\ 51.15\\ 60.85\\ 64.93\\ 67.96\\ 70.32\end{tabular}} & \begin{tabular}[c]{@{}l@{}}36.64\\ 51.14\\ \underline{62.81}\\ \underline{67.05}\\ \underline{70.30}\\ \underline{73.55}\end{tabular} \\ \hline
                                                                                &                                                                                                                                                &                                                                                                                  &                                                                                                            &                                                                                                            &                                                                                                            &                                                                                                            &                                                                                       \\ \hline
\multicolumn{1}{|c|}{\begin{tabular}[c]{@{}c@{}}BLIP-2\\ Retriever\end{tabular}} & \multicolumn{1}{c|}{\begin{tabular}[c]{@{}c@{}}R@20: 36.87\\ R@40: 46.45\\ R@60: 52.06\\ R@80: 55.98\\ R@100:58.91\\ R@120:61.30\end{tabular}} & \multicolumn{1}{c|}{\begin{tabular}[c]{@{}c@{}}Turn 0\\ Turn 1\\ Turn 2\\ Turn 3\\ Turn 4\\ Turn 5\end{tabular}} & \multicolumn{1}{c|}{\begin{tabular}[c]{@{}c@{}}36.87\\ \underline{55.71}\\ 58.87\\ 59.67\\ 59.88\\ 59.92\end{tabular}} & \multicolumn{1}{c|}{\begin{tabular}[c]{@{}c@{}}36.87\\ 51.92\\ 58.98\\ 63.64\\ 66.65\\ 69.33\end{tabular}} & \multicolumn{1}{c|}{\begin{tabular}[c]{@{}c@{}}36.87\\ 51.53\\ 56.54\\ 59.79\\ 62.23\\ 64.29\end{tabular}} & \multicolumn{1}{c|}{\begin{tabular}[c]{@{}c@{}}36.87\\ 51.73\\ 61.06\\ 65.78\\ 68.32\\ 70.50\end{tabular}} & \begin{tabular}[c]{@{}c@{}}36.87\\ 51.88\\ \underline{61.67}\\ \underline{67.51}\\ \underline{70.37}\\ \underline{73.82}\end{tabular} \\ \hline
\end{tabular}
}
\label{tab:table2}
\vspace{-0.4cm}
\end{table}

% \vspace{0.1cm}
% \noindent\textbf{4.5 The Impact of Different LLMs}
\subsection{The Impact of Different LLMs}

\begin{figure}[t!]
   \includegraphics[width=\linewidth]{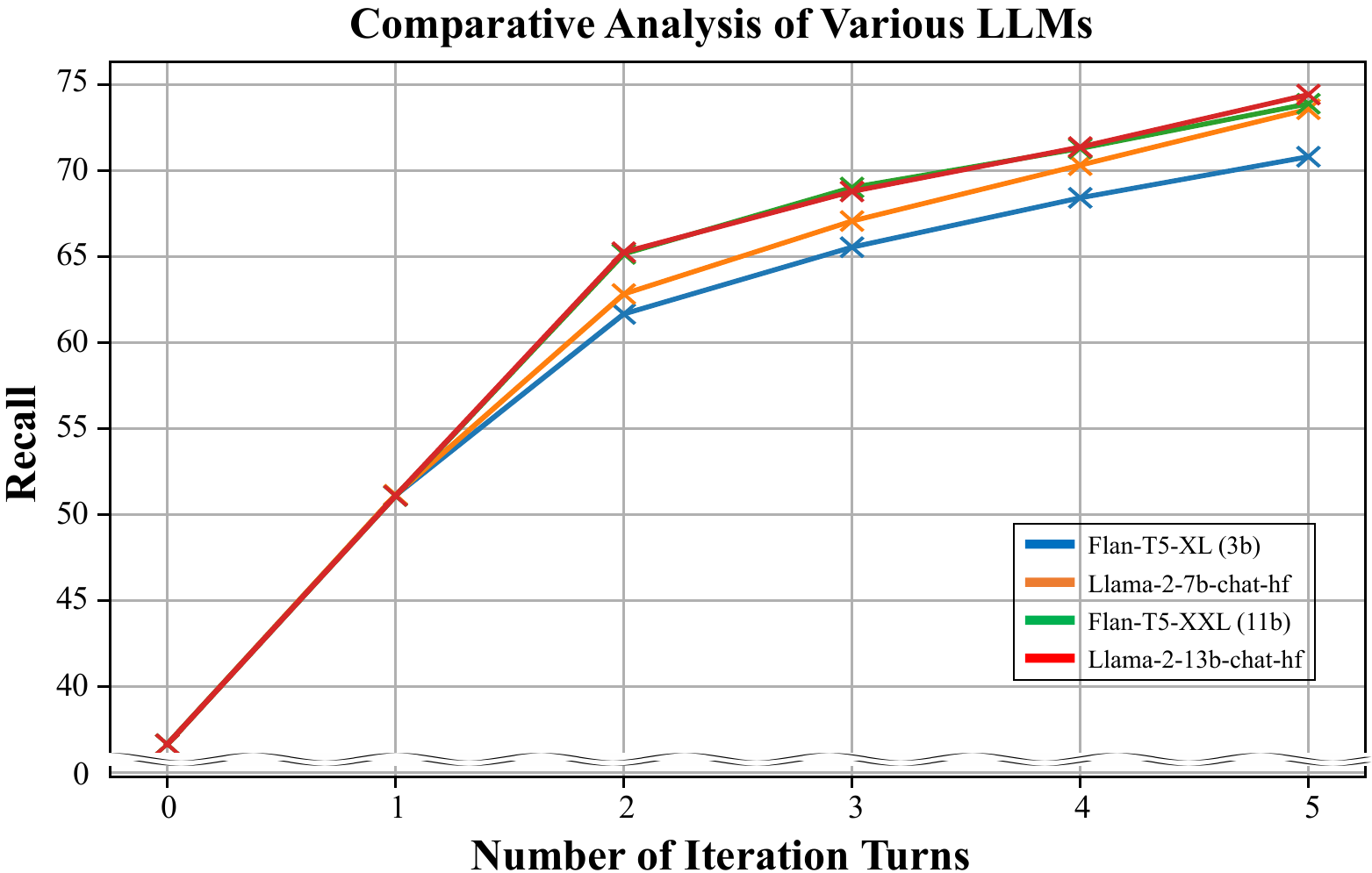}
  \vspace{-0.8cm}
  \caption{The performance evaluation of systems utilizing two variants of LLMs and four different sizes for CoT query summaries. The CLIP model is employed as the image retriever in this setup.
  % The performance of the systems with two types of LLMs and four sizes for CoT query summary. CLIP model is applied as the image retriever.
  }
  \label{fig:figure ablation}
  \vspace{-0.5cm}
\end{figure}

\begin{figure}[t!]
   \includegraphics[width=\linewidth]{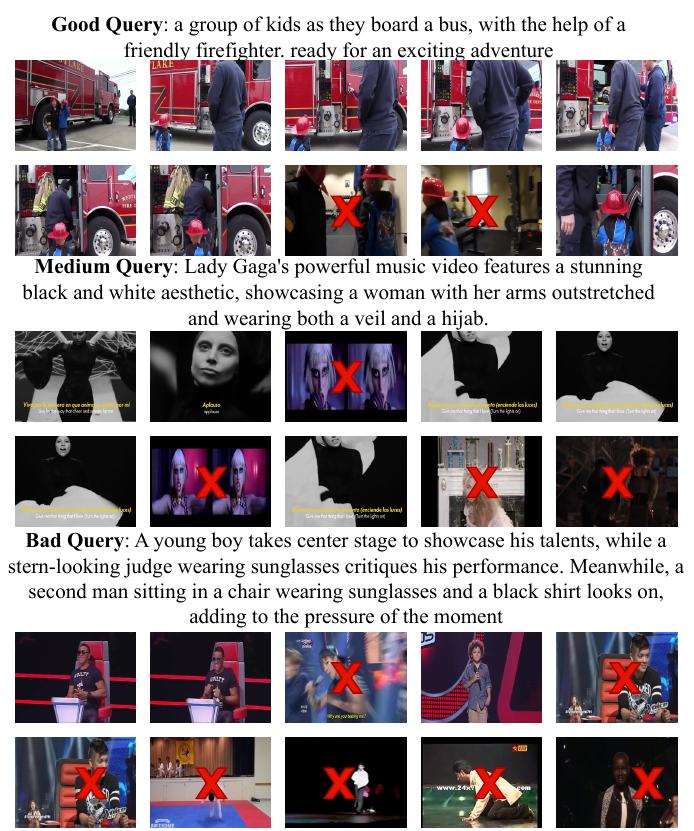}
  \vspace{-0.6cm}
  \caption{Example queries and search results from the proposed system with CoT query summaries are showcased. Irrelevant images are denoted with a red cross for clarity.
  % Example queries and search results from the proposed system with CoT query summary. Different quality of queries are presented. The irrelevant images are labeled with a red cross.
  }
  \label{fig:figure example}
  \vspace{-0.5cm}
\end{figure}

In order to assess the practical capabilities and constraints of the LLM-based query editing technique, we juxtapose two sets of models of varying sizes, as depicted in Figure \ref{fig:figure ablation}. We have chosen two sizes of the FLAN-T5 model~\cite{Chung2022ScalingIL} (Flan-T5-XL and Flan-T5-XXL) and two sizes of the LLama-2 model~\cite{Touvron2023Llama2O} (Llama-2-7b-chat-hf and Llama-2-13b-chat-hf), with respective model sizes of 3B, 11B, 7B, and 13B. These models are applied to the CoT query summary in our interactive image retrieval system, which employs a CLIP retriever. The decision to refrain from utilizing larger models is attributed to the constraints of our computational resources. Furthermore, the employment of excessively large LLMs would considerably diminish the efficiency of the interactive system, rendering it impractical for real-world applications. From Figure \ref{fig:figure ablation}, it is observed that all the tested LLMs exhibit similar performance. The largest LLM, LLama-2-13b-chat-hf, yields a 4\% recall improvement throughout 5 interaction turns compared to the smallest model, Flan-T5-XL. Interestingly, the smallest LLM with a CoT prompt still outperforms the largest LLM when other prompt methods are used. This further attests to the effectiveness of our CoT query summary method.    
\label{sec:ablation}

\section{Conclusion and Future Work}
% \vspace{-0.1cm}

% Jiahong's version
In summary, this paper presents a novel interactive image retrieval system aimed at overcoming the limitations of traditional single-turn methods. Through the incorporation of multi-turn interactions and continuous refinement of queries based on user relevance feedback, our system significantly improves the quality of text-based queries in natural language space. We introduce an LLM-based denoiser to refine text-based query expansions, effectively addressing inaccuracies and enhancing specificity in image descriptions generated by captioning models. We curated a meticulously designed evaluation dataset, adapted from the MSR-VTT video retrieval dataset, to offer multiple relevant ground truth images for each query, thus mitigating limitations in existing datasets. Through extensive experiments, our proposed system demonstrates a substantial 10\% improvement in recall over baseline methods after 6 turns of iterations, achieving state-of-the-art performance. These contributions advance the field of interactive image retrieval and provide valuable insights for future research in this domain.
Future research could explore integrating user preferences and domain-specific knowledge to further enhance the accuracy of interactive image retrieval systems. Leveraging multi-turn interaction, VLMs, and LLMs presents an innovative approach that holds promise for developing more user-centric and adaptable image retrieval systems.

% Hongyi's version
% In conclusion, this research addresses the limitations of existing single-run image retrieval methods by introducing an innovative interactive image retrieval system. Recognizing the potential inaccuracies and noise introduced by image captions, the system employs a multi-turn approach, integrating image captioning and leveraging LLMs for denoising. The proposed denoiser enhances the quality of text-based queries, refining them before each iteration of the retrieval model. 
% The experiments conducted on MSR-VTT dataset demonstrate the effectiveness of the system, particularly the 10\% recall improvement after 6 interaction turns compared to single-turn retrieval and vector space-based models. The contributions of this work extend to the development of a user-friendly, multi-turn text-to-image retrieval system, the application of Visual Language Models for query expansion, and the utilization of LLMs for denoising to improve system performance. Overall, these contributions highlight the potential of the proposed approach to enhance the efficiency and effectiveness of image retrieval systems in real-world scenarios. Future work can explore incorporating user preferences and domain-specific knowledge for enhanced accuracy. This novel approach, leveraging multi-turn interaction, VLMs, and LLMs, paves the way for more user-centric and adaptable image retrieval systems.

\pagebreak

% The next two lines define the bibliography style to be used, and the bibliography file.
\bibliographystyle{ACM-Reference-Format}
\bibliography{sample-base}

% 
% If your work has an appendix, this is the place to put it.
\appendix

\end{document}